\documentclass[twocolumn]{aastex62}

\usepackage{amsmath}
\usepackage[T1]{fontenc}
\usepackage{apjfonts} 

\newcommand{\code}[1]{\texttt{#1}}
\newcommand{\mesa}{\code{MESA}}
\newcommand{\MESA}{\mesa}

\newlength{\apjcolwidth}
\newlength{\figwidth}
\newlength{\doublewide}
\begin{document}
\title{
Increases to Inferred Rates of Planetesimal Accretion Due to
Thermohaline Mixing in Metal Accreting White Dwarfs
}

\author[0000-0002-4791-6724]{Evan B. Bauer}
\affiliation{Department of Physics, University of California, Santa Barbara, CA 93106, USA}
\correspondingauthor{Evan B. Bauer}
\email{ebauer@physics.ucsb.edu}

\author{Lars Bildsten}
\affiliation{Department of Physics, University of California, Santa Barbara, CA 93106, USA}
\affiliation{Kavli Institute for Theoretical Physics, University of California, Santa Barbara, CA 93106, USA}

\begin{abstract}
Many isolated, old white dwarfs (WDs) show surprising evidence of
metals in their photospheres. Given that the timescale for
gravitational sedimentation is astronomically short, this is taken as
evidence for ongoing accretion, likely of tidally disrupted
planetesimals. The rate of such accretion, ${\dot M_{\rm acc}}$, is
important to constrain, and most modeling of this process relies on
assuming an equilibrium between diffusive sedimentation and metal
accretion supplied to the WD's surface convective envelope. Building
on earlier work of Deal and collaborators, we show that high ${\dot
M_{\rm acc}}$ models with only diffusive sedimentation are unstable
to thermohaline mixing and that models which account for the enhanced
mixing from the active thermohaline instability require larger
accretion rates, sometimes reaching ${\dot M_{\rm acc} \approx 10^{13}
  \, \rm g \, s^{-1}}$ to explain observed Calcium abundances.
We present results from a grid of \texttt{MESA} models that
include both diffusion and thermohaline mixing. These results demonstrate
that both mechanisms are essential for understanding metal pollution
across the range of  polluted WDs with hydrogen atmospheres.
Another consequence of active thermohaline mixing is that the observed
metal abundance ratios are identical to accreted material.
\end{abstract}


\section{Introduction}

Despite the short sedimentation timescales for metals that should lead
to pure hydrogen atmospheres, a large fraction ($25-50\%$) of DA white
dwarfs (WDs) show signatures of atmospheric metal pollution \citep{Zuckerman03}.
Though radiative levitation can prevent some element
settling in WDs with $T_{\rm eff}\gtrsim 20,000\,\rm K$
\citep{Chayer95a,Chayer95b,Chayer14}, more than $25\%$ of WDs can
only be explained with ongoing accretion providing a continued supply
of metals to the surface \citep{Koester14}.
Many of these WDs are thought to be actively accreting debris from disrupted
planetesimals that are perturbed within the WD tidal radius, and some
show anomalous infrared emission indicative of an accreting disk
\citep{Jura03,Farihi09,Girven12,Vanderburg15,Farihi16}.
Detailed atmosphere models for mixing and
metal sedimentation allow for inferences of accretion rates and
compositions \citep{Vauclair79,Dupuis92,Dupuis93,Koester06,Koester09},
and thus these objects serve as unique laboratories for observing the
interior bulk composition of planetesimals around WDs
\citep{Zuckerman07,Boris12,Dufour12,Koester14,Jura14}.

Much recent work in this field has relied on the assumption of
equilibrium between accretion and diffusive sedimentation. The
large mean molecular weight of accreted material compared to the
hydrogen atmosphere can lead to additional mixing due to the thermohaline
instability \citep{Deal13,Wachlin17}, though see \cite{Koester15} for
a critique of its efficacy. We perform
time-dependent stellar evolution calculations using \mesa\
\citep{Paxton11,Paxton13,Paxton15,Paxton18}.
Our models that account for
diffusion and thermohaline mixing indicate that both mechanisms are
essential for understanding the range of observed
parameters for polluted WDs. Our work spans a large
range of effective temperatures and accretion rates, allowing new
accretion inferences for hydrogen atmosphere WDs with $T_{\rm
  eff}\lesssim 20,000\,\rm K$.

\begin{figure*}
\begin{center}
\includegraphics[width=\apjcolwidth]{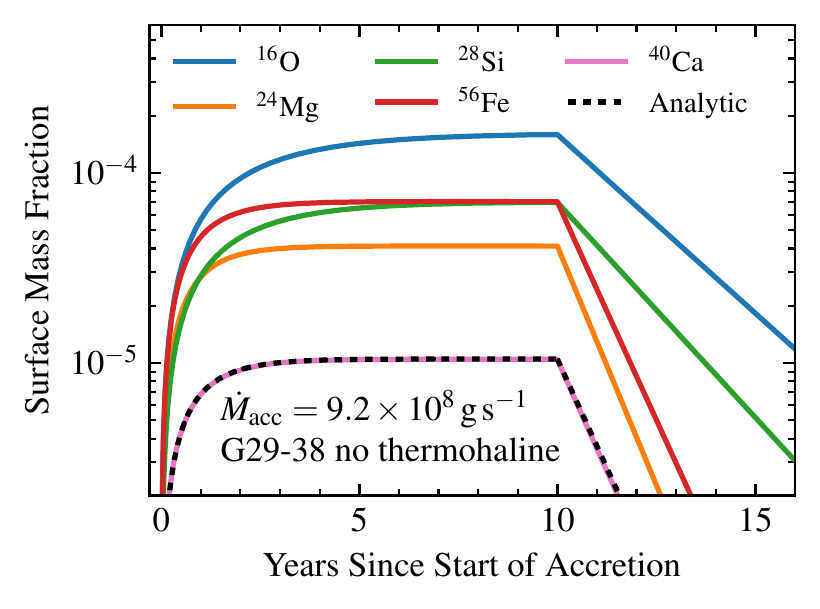}
\hspace{2em}
\includegraphics[width=\apjcolwidth]{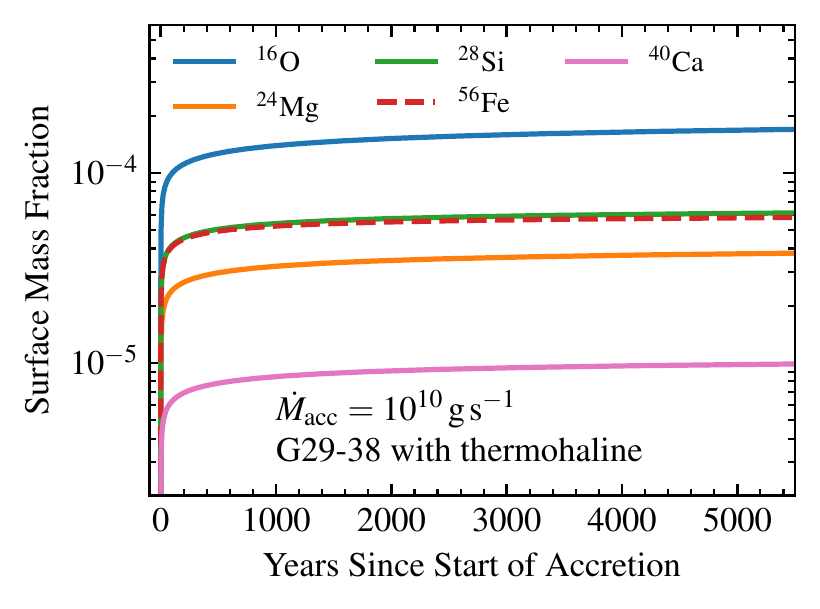}
\caption{
{\it Left:} \mesa\ model of G29-38 matching observed abundances
\citep{Xu14} after accreting earth-like composition at a total rate of
${\dot M_{\rm acc}=9.2\times 10^8\,\rm g\, s^{-1}}$
for 10 years to approach equilibrium surface abundances. 
Accretion ceases after 10 years to illustrate the exponential decay of
observable pollution governed by diffusion in the absence of
accretion. No thermohaline mixing is included in this model.
{\it Right:}~\mesa\ model of G29-38 including thermohaline mixing
accreting metals in the ratios observed at the WD surface at
${\dot M_{\rm acc}=10^{10}\,\rm g\, s^{-1}}$
for 5,000 years to approach observed surface abundances.
}
\label{fig:episode}
\end{center}
\end{figure*}

In Section~\ref{sec.diffusion}, we discuss standard diffusion
calculations and their relation to sedimentation in \mesa\ models.
In Section~\ref{sec.instability}, we examine conditions for thermohaline
instability in layers containing accreted metals and explore the
impact of thermohaline mixing in our \mesa\ models.
In Section~\ref{sec.impact}, we present the resulting
implications for observed polluted WDs with hydrogen
atmospheres. We find that WDs with $T_{\rm eff}\gtrsim 10,000\,\rm K$
require accretion rates several orders of
magnitude larger than previously inferred, with the largest as high as
${\dot M_{\rm acc}\approx 10^{13}\,\rm g\, s^{-1}}$ when
accounting for thermohaline mixing.
In Section~\ref{sec.conclusion}, we discuss avenues for extending and
refining the grid of models.

\section{Gravitational Sedimentation}
\label{sec.diffusion}

In accretion-diffusion equilibrium, the observed metal abundances in
the outer convective zone of a polluted WD are simply related to those
in the accreted material. The timescale for convective mixing is
rapid, so accreted material is quickly 
distributed throughout the convection zone. The observable mass fraction of
an accreted pollutant $X_{{\rm cvz},i}$ is then related to the
accretion rate $\dot M_i$ for that pollutant by
(cf.~\citealt{Vauclair79, Dupuis93, Koester09})
\begin{equation}
M_{\rm cvz}\frac{dX_{{\rm cvz},i}}{dt}=\dot M_i-
\frac{X_{{\rm cvz},i}M_{\rm cvz}}{\tau_{{\rm diff},i}}~,
\label{eq:differentialX}
\end{equation}
where the sedimentation time for element $i$ is
\begin{equation}
\tau_{{\rm diff},i}\equiv\frac{M_{\rm cvz}}{4\pi r^2\rho v_{{\rm diff},i}}~,
\end{equation}
and $v_{{\rm diff},i}$ is the downward sedimentation velocity of the accreted pollutant
at the base of the surface convection zone, at density $\rho$ and
radius $r$,  where it sinks away from the fully mixed surface region.

Assuming $\tau_{{\rm diff},i}$, $\dot M_i$, and $M_{\rm cvz}$ are time
independent, Equation~\eqref{eq:differentialX} gives
\begin{equation}
X_{{\rm cvz},i}(t)=X_{{\rm cvz},i}(0)\, e^{-t/\tau_{{\rm diff},i}}
+\frac{\dot M_i}{M_{\rm cvz}}\tau_{{\rm diff},i} 
\left(1-e^{-t/\tau_{{\rm diff},i}}\right)~,
\label{eq:analytic}
\end{equation}
and for $t\gg\tau_{{\rm diff},i}$ the mass fraction approaches the
equilibrium value
\begin{equation}
X_{{\rm eq},i}=\frac{\dot M_i}{M_{\rm cvz}}\tau_{{\rm diff},i}~.
\label{eq:tau_eq}
\end{equation}
Since sedimentation timescales in hydrogen WD atmospheres are short
($\tau_{\rm diff}\sim {\rm days}-10^4\, {\rm years}$, \citealt{Koester09}),
Equation~\eqref{eq:tau_eq} is typically used to infer elemental accretion rates
from observed abundances assuming $X_{{\rm obs},i}=X_{{\rm eq},i}$. A total
accretion rate can be found simply by adding the individual
contributions of each observed pollutant ($\dot M_{\rm acc}\equiv\sum_i\dot M_i$), 
or by scaling to a fiducial composition when other
elements are not directly observed but are expected to be present
(e.g.~$\dot M_{\rm acc}=\dot M_{\rm Ca}/X_{\rm acc,Ca}$).

The theoretical ingredients for the preceding calculation
are diffusion timescales and surface convection zone masses
\citep{Koester06,Koester09}. We calculate
these as part of time-dependent WD evolutionary models with
hydrogen atmospheres using \mesa\ version 10398.
In particular, our treatment of diffusion is based on a complete
time-dependent solution of the Burgers equations for diffusion
\citep{Burgers69}, adapted to be appropriate for any degree of electron
degeneracy in WDs as described in detail in \cite{Paxton18}. This
treatment yields diffusion velocities for each species in the plasma
everywhere in the stellar model.
Our results for diffusion timescales and convection zone masses in
WDs with hydrogen atmospheres are comparable to those of 
\cite{Koester09}.\footnote{Most recent tables found at
\url{http://www1.astrophysik.uni-kiel.de/~koester/astrophysics/astrophysics.html}.}

The convection prescription for our models is ML2 \citep{Bohm71} with $\alpha_{\rm MLT}=0.8$.
Our settings for surface boundary conditions rely on either
a grey iterative procedure ($T_{\rm eff}>9,000\,\rm K$) described in
\cite{Paxton13}, or the WD atmosphere tables in \mesa\
($T_{\rm eff}<9,000\,\rm K$),
which are adapted from \cite{Rohrmann12}.
Diffusion coefficients are those of \cite{Stanton16} as implemented in
\cite{Paxton18}, which produce comparable results to those of
\cite{Paquette86coeff}. 
More details are presented in a forthcoming paper. For WDs
with no surface convection zone (${T_{\rm eff}\gtrsim 15,000\, \rm
  K}$), we take the surface region in which
to evaluate $X_{\rm cvz}$ and $M_{\rm cvz}$ to be everywhere above the
photosphere in the model (optical depth $\tau_{\rm Ross}=2/3$),
with $\tau_{\rm diff}$ evaluated at the photosphere.
See \cite{Boris12} for a thorough discussion justifying this choice.

The left panel of Figure~\ref{fig:episode} shows that a \mesa\ model
of an accretion episode with constant $\dot M_{\rm acc}=9.2\times
10^8\,\rm g\, s^{-1}$ 
(accreted mass fractions ${X_{\rm Fe}=0.307}$, ${X_{\rm O}=0.295}$, 
${X_{\rm Mg}=0.199}$, $X_{\rm Si}=0.153$, $X_{\rm Ca}=0.046$)
agrees with the prediction of
Equation~\eqref{eq:analytic} when thermohaline mixing is not
considered. This model is tuned to match the observed properties of
\mbox{G29-38} presented by \cite{Xu14}, including $T_{\rm eff}=11,820\,\rm K$,
$\log g=8.4$, and the abundances presented in their
Table~3. Our \mesa\ model has diffusion timescales on the order of
1-2 years,  roughly a factor of 5 longer than those reported by
\cite{Xu14} due to a larger surface convection zone in the \mesa\
model ($M_{\rm cvz}=6\times 10^{-14}\, M_\odot$).
This difference arises because convection zone depths are very
sensitive to $T_{\rm eff}$ around the temperature for G29-38, and our
models show growth of the surface convection zone slightly sooner as
the WD cools compared to those of \cite{Koester09} in this regime.
Hence, we infer an accretion rate that is
$40\%$ larger than the value \cite{Xu14} report, but we find excellent
agreement with their relative diffusion timescales and composition of
the accreted material.

\section{Thermohaline Instability}
\label{sec.instability}

\cite{Deal13} and \cite{Wachlin17} have noted
that thermohaline mixing may significantly alter the accretion rates
inferred from observed abundances in polluted WDs.
\cite{Koester15} argued against the efficacy of this instability for a
few cases. Here we show the regime in which it operates and its
overall impact, which can be large.

\subsection{Onset of the Instability Beneath the Convection Zone}

Thermohaline instability can occur when fluid is stable to convection
according to the Ledoux criterion, but has an inverted mean molecular
weight gradient:
\begin{equation}
\nabla_T-\nabla_{\rm ad}< \frac{\varphi}{\delta} \nabla_\mu <0~,
\label{eq:ineq1}
\end{equation}
where $\nabla_T=(\partial\ln T/\partial\ln P)$ is the
temperature gradient, ${\nabla_{\rm ad}=(\partial\ln T/\partial\ln
  P)_s}$ is the adiabatic temperature gradient, 
$\nabla_\mu=(\partial\ln\mu/\partial\ln P)$ is the mean molecular
weight gradient,
${\varphi=(\partial\ln\rho/\partial\ln\mu)_{P,T}}$, and
${\delta=-(\partial\ln\rho/\partial\ln T)_{P,\mu}}$. A fluid satisfying
Equation~\ref{eq:ineq1} alone is not guaranteed to experience
thermohaline instability; the double-diffusive nature of the
instability requires that microscopic particle transport between fluid
elements be slow compared to thermal transport so that perturbed
elements maintain their composition long enough for the instability to
grow. This extra condition is \citep{Baines69,Brown13,Garaud18}:
\begin{equation}
\frac{(\varphi/\delta)\nabla_\mu}{\nabla_T-\nabla_{\rm ad}}>\frac{\kappa_\mu}{\kappa_T}~,
\label{eq:ineq2}
\end{equation}
where $\kappa_\mu$ and $\kappa_T$ are the particle and thermal diffusivities.

\begin{figure}
\begin{center}
\includegraphics[width=\apjcolwidth]{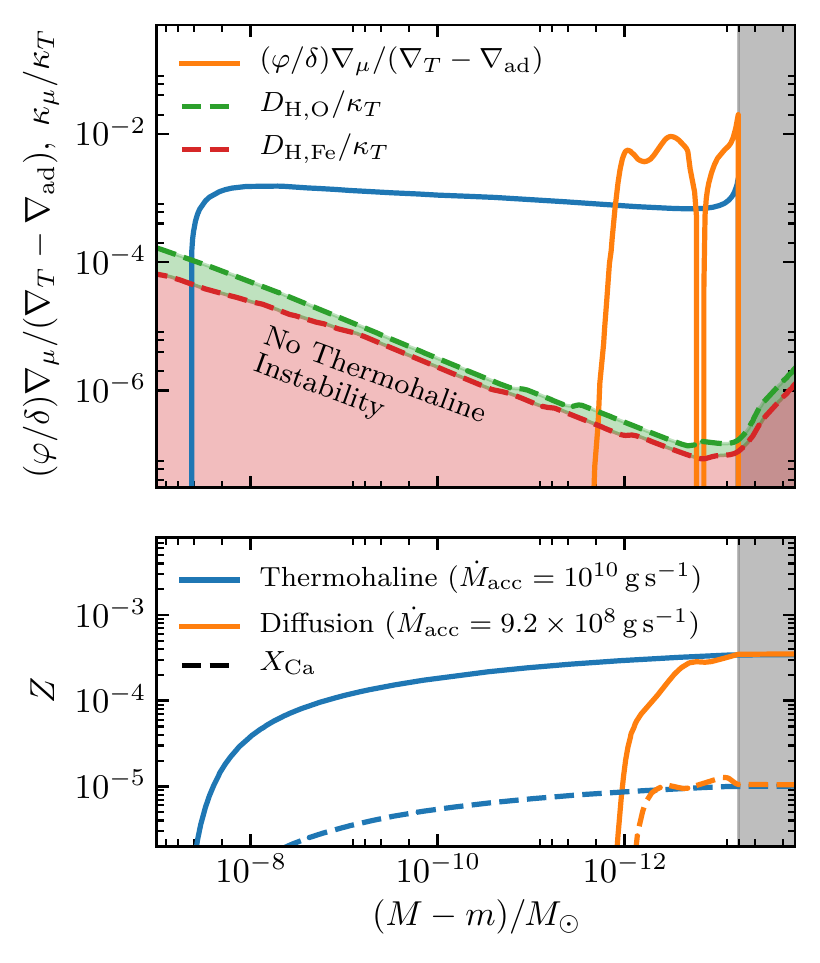}
\caption{
Profiles of quantities relevant to the onset of thermohaline
instability in  G29-38 models.
The lower panel shows the metallicity profile 
with (blue) and without (orange) thermohaline mixing enabled according
to Equation~\ref{eq:KRTD} with $\alpha_{\rm th}=1$, for the same
models as shown in Figure~\ref{fig:episode}.
The upper panel shows the quantities necessary for evaluating
the inequality given by Equation~\eqref{eq:ineq2} in both cases,
confirming that these profiles are thermohaline unstable.
The gray shaded region represents the convective envelope.
}
\label{fig:thermo_abund}
\end{center}
\end{figure}

The mean molecular weight gradient is inverted in the radiative zone
beneath the outer convective layer of these polluted WDs.
Heat is carried there via radiative diffusion, so
\begin{equation}
\kappa_T=\frac{4acT^3}{3\kappa\rho^2C_P}~.
\end{equation}
We use the diffusion coefficients of oxygen
or iron in hydrogen as representative of the particle diffusivity
relevant for the mean molecular weight of a fluid element:
\begin{equation}
\kappa_\mu\approx D_{\rm H,O}\text{ or }D_{\rm H,Fe}~.
\end{equation}
We obtain these coefficients from the same routines based on
\cite{Stanton16} that we use to calculate coefficients for element
diffusion \citep{Paxton18}.

Figure~\ref{fig:thermo_abund} shows the quantities relevant for evaluating
the criterion for thermohaline instability given in
Equation~\ref{eq:ineq2} for the G29-38 \mesa\ model.
The metallicity ($Z$) profile for the model is unstable according to
Equation~\ref{eq:ineq2}, and the result of enabling thermohaline
mixing will be to mix a significant amount of the metals deeper into
the star. 
A significantly larger accretion rate is then required to match the
observed surface pollution in G29-38,
similar to the results of \cite{Wachlin17}.

\subsection{Outcomes when Thermohaline Mixing is Included}

The default thermohaline mixing treatment in \mesa\ \citep{Paxton13}
follows \cite{KRT80} in defining the mixing coefficient:
\begin{equation}
D_{\rm th}=\alpha_{\rm th}\kappa_T\frac{3}{2}\frac{(\varphi/\delta)\nabla_\mu}{\nabla_T-\nabla_{\rm ad}}~,
\label{eq:KRTD}
\end{equation}
where $\alpha_{\rm th}$ is a dimensionless efficiency parameter that
should be near unity.
\mesa\ also offers options for using more recent
treatments of thermohaline mixing due to \cite{Traxler11} and
\cite{Brown13}, which attempt to constrain free parameters by
calibrating effective 1D prescriptions to 3D simulations.

\begin{figure} 
\begin{center}
\includegraphics[width=\apjcolwidth]{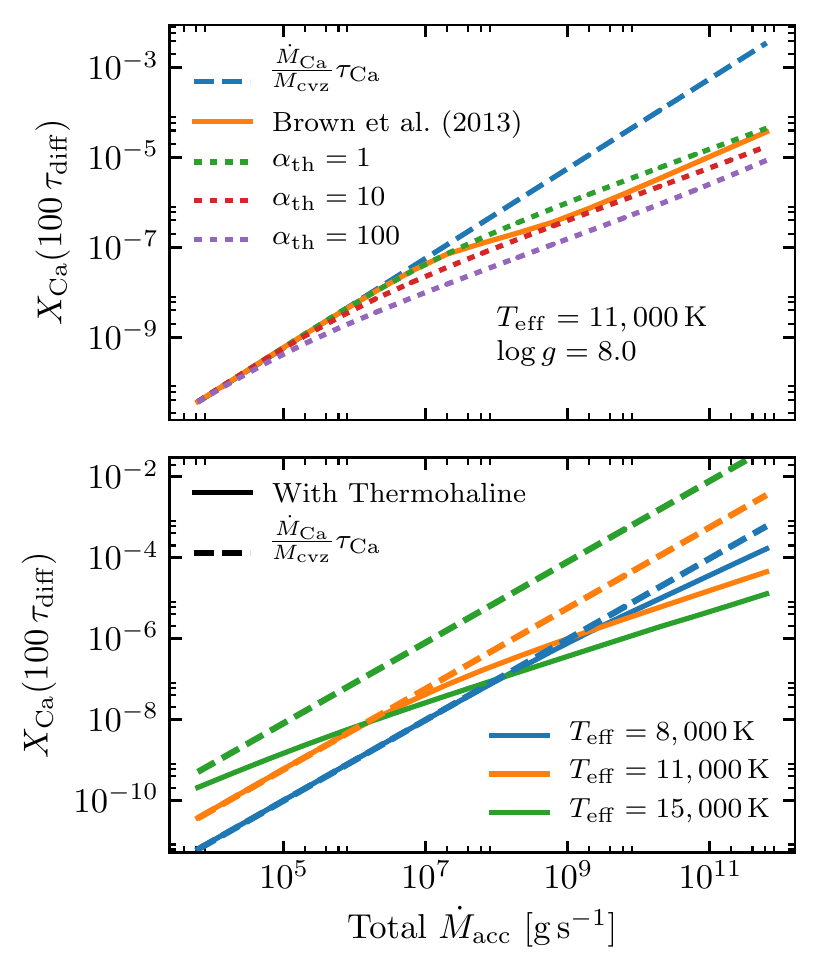}
\caption{
Surface Ca mass fraction after 100 diffusion timescales
as a function of total metal accretion rate
for a $0.6\, M_\odot$ ($\log g=8.0$) WD
models accreting metals in bulk earth ratios. The upper panel shows
several thermohaline prescriptions for a $T_{\rm eff}=11,000\,\rm
K$ WD. The lower panel shows the WD at several temperatures for the
thermohaline prescription of Equation~\ref{eq:KRTD} with $\alpha_{\rm th}=1$.
}
\label{fig:thermo_eff}
\end{center}
\end{figure}

Figure~\ref{fig:thermo_eff} shows the surface Ca mass fraction in
polluted $0.6\, M_\odot$ ($\log g=8.0$) WD  models after accreting
bulk earth material \citep{McD} at $\dot M_{\rm acc}\in
(10^4,10^{12})\,\rm g\, s^{-1}$ for many diffusion timescales. These
\mesa\ models include element diffusion at all times.
The upper panel includes models with thermohaline mixing according to
Equation~\ref{eq:KRTD} (several values of $\alpha_{\rm th}$) and also
according to \cite{Brown13}. These results suggest that
Equation~\ref{eq:KRTD} reasonably captures the net effects of thermohaline
mixing on pollution for efficiencies in the range ${1<\alpha_{\rm th}<10}$,
though note that \cite{Vauclair12} have argued that more
mixing should occur near the limit of instability to convection
(${\nabla_\mu \sim \nabla_{T} - \nabla_{\rm ad}}$).
We elect to employ the treatment of Equation~\ref{eq:KRTD} with
$\alpha_{\rm th} = 1$ for the models presented in the remainder of
this work. Our choice of mixing prescription thus
represents a reasonable but conservative estimate of the magnitude
and total impact of thermohaline mixing for pollution.

The lower panel of Figure~\ref{fig:thermo_eff} shows the Ca pollution for models
at several different values of $T_{\rm eff}$ using the thermohaline
prescription of Equation~\ref{eq:KRTD} with $\alpha_{\rm th}=1$.
Models with thin surface
convection zones ($T_{\rm eff}=11,000\,\rm K; 15,000\,\rm K$) 
experience significant dilution of surface metals due to thermohaline
mixing, while models with larger surface convection zones ($T_{\rm
  eff}=8,000\,\rm K$) distribute accreted metals to
an extent that $\nabla_\mu$ is rarely large enough to drive
significant thermohaline mixing beneath the convection zone.
In this case, diffusive
sedimentation governs the timescale for settling and observed
equilibrium abundances.

Unlike models with only element diffusion, those with thermohaline
mixing do not approach a true equilibrium composition on a short
timescale. Instead, they approach a quasi-steady state
composition near the surface for timescales represented in
Figure~\ref{fig:thermo_eff} ($\sim 100\, \tau_{\rm diff}$),
but these quasi-steady abundances may vary by
factors of a few if accretion continues over very long timescales
($\sim$Myr) as the thermohaline mixing region continues to extend
deeper into the hydrogen envelope.
Thermohaline mixing ceases to extend inward only when it encounters
the diffusive tail of helium near the base of the hydrogen
envelope. The $\nabla_\mu$ from this helium tail counteracts that
from metals mixing inward, preventing any further thermohaline
instability. For the hydrogen envelopes in our $M=0.6\, M_\odot$ WD
models ($M_{\rm H}\approx 10^{-6}\, M_\odot$), the thermohaline
mixing encounters the helium layer only after long periods ($\sim$Myr)
of sustained high accretion rates ($\dot M_{\rm acc}\gtrsim 10^{10}
\,\rm g\, s^{-1}$). Hence, this effect is not significant for the
models we present in this work, but it may be important for WDs with
much thinner hydrogen envelopes ($M_{\rm H}\lesssim 10^{-8}\,
M_\odot$), where it could lead to higher observed surface pollution
by preventing thermohaline mixing that would otherwise occur.

\section{Accretion Rates and Compositions}
\label{sec.impact}

\begin{figure} 
\begin{center}
\includegraphics[width=\apjcolwidth]{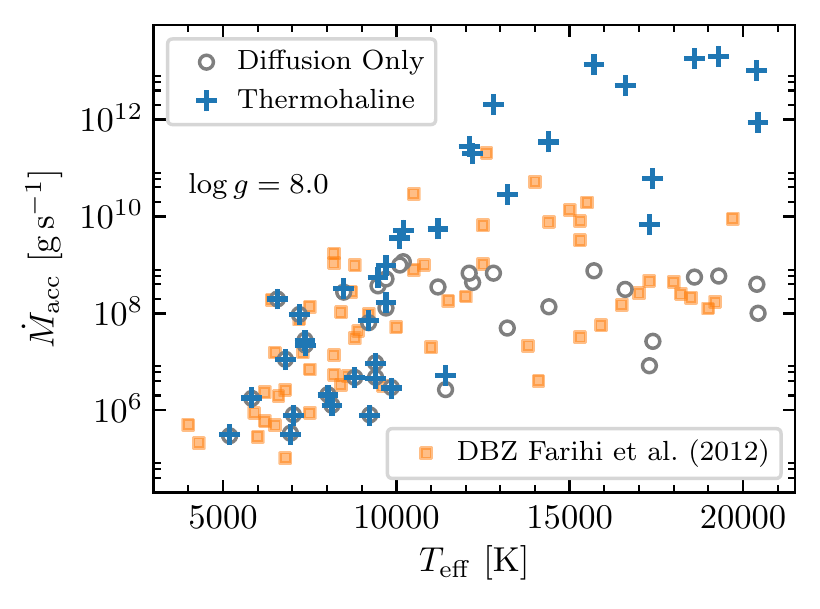}
\caption{
Accretion rates inferred with (blue crosses) and without (open circles)
thermohaline mixing from the observed Ca abundances for the 38 WDs
given in Table~1 of \cite{Koester06}, assuming $\log g=8.0$ and 
bulk earth composition in the accreting material.
The orange points show the rates inferred for He
atmosphere WDs by \cite{Farihi12} for comparison.
}
\label{fig:thermo_mdot}
\end{center}
\end{figure}

In order to show the impact of our work on inferring
accretion rates, we built an interpolating tool to map observed Ca
abundances to total metal accretion rates for \mesa\ runs that
include thermohaline mixing and diffusion.
The \mesa\ runs used for this interpolation consist of 12 different
WD models ($M=0.6\, M_\odot$, $\log g=8.0$) 
in the temperature range ${T_{\rm eff}\in (6000,20500) \,\rm K}$,
each accreting bulk earth composition \citep{McD} for 100 diffusion timescales at
17 rates in the range $\dot M_{\rm acc}\in (10^4,10^{12})\,\rm g\, s^{-1}$, 
with diffusion enabled along with thermohaline mixing
according to Equation~\ref{eq:KRTD} with $\alpha_{\rm th}=1$.
At fixed $T_{\rm eff}$, a given observed Ca abundance corresponds to a
unique value of total $\dot M_{\rm acc}$ (as seen in the lines in the
lower panel of Figure~\ref{fig:thermo_eff}), and we interpolate between
results from \mesa\ runs at different $T_{\rm eff}$ to yield a result
for $\dot M_{\rm acc}$ as a function of $T_{\rm eff}$ and observed Ca
abundance.
Figure~\ref{fig:thermo_mdot} shows the results of this
procedure for inferring total $\dot M_{\rm acc}$ for the 38 polluted
hydrogen atmosphere WDs given in Table~1 of \cite{Koester06}.

Figure~\ref{fig:thermo_mdot} also compares against
\mesa\ calculations assuming only diffusion governs the equilibrium state,
where we use Equation~\ref{eq:tau_eq} to infer $\dot M_{\rm acc}$, and
assume bulk earth abundances for the accreted material. 
For objects with ${T_{\rm eff}>10,000\,\rm K}$, the inferred accretion
rate can increase by several orders of magnitude when accounting
for thermohaline mixing, confirming the earlier work of
\cite{Deal13} and \cite{Wachlin17}.

Our models assume that the observed Ca is only a fraction
of the total metals being accreted. This is important because all accreted metals
participate in determining the $\nabla_\mu$ that
drives thermohaline mixing. This should be justified, since Ca is often
the most easily identifiable element
in spectra even when other metals are present, and for
the objects in which many metals are identified, compositions appear
roughly consistent with bulk earth \citep{Boris12,Jura14,Xu14}.
Our inferences are also currently limited by the fact that all input
models have a mass of $0.6\, M_\odot$ ($\log g=8.0$), which is not
always consistent with the observed WDs. More \mesa\ runs are
necessary to build a tool that can interpolate over $\log g$ as well
as $T_{\rm eff}$. However, we do not expect the small corrections due to
different $\log g$'s to lead to qualitative changes in the several
orders of magnitude effect seen in Figure~\ref{fig:thermo_mdot}.

For objects where multiple pollutants are observed, it
should be possible to estimate relative accreted abundances, but there
are two distinct regimes. For low accretion rates or thick
convective envelopes where the thermohaline instability is not
excited, the equilibrium state of each element is separately governed
by Equation~\ref{eq:tau_eq}, so that the inferred relative abundance
is related to observations by
\begin{equation}
\frac{X_{{\rm obs},1}}{X_{{\rm obs},2}} 
=\frac{\tau_{{\rm diff},1}X_{{\rm acc},1}}{\tau_{{\rm diff},2}X_{{\rm acc},2}}~.
\end{equation}
However, when thermohaline mixing
dominates over diffusion, the diffusion timescales play no
role. Instead the mixing coefficient $D_{\rm th}$ applies equally to
all elements, resulting in the observed relative abundances of metals
matching the ratios in the accreted material.

Returning to \mesa\ models of G29-38, when thermohaline mixing is
included, we achieve a good match to observed pollution with an
accretion rate of ${\dot M_{\rm acc}\gtrsim 10^{10}\,\rm g\, s^{-1}}$
(depending on thermohaline efficiency), higher than the value
${\dot M_{\rm acc}\approx 3\times 10^9\,\rm g\, s^{-1}}$
suggested by \cite{Wachlin17}.
The significant thermohaline mixing also means
that the best match to observed composition is achieved with a model
that accretes metals in the same ratios as those observed at the
photosphere, with no correction for relative diffusion
timescales. Whereas the model without thermohaline mixing matches bulk
earth composition remarkably well, models of G29-38 with thermohaline mixing
imply a significantly more oxygen rich composition of the accreted
material.

\section{Conclusions and Future Work}
\label{sec.conclusion}

Our \mesa\ models indicate that inferred accretion rates in polluted
WDs with hydrogen dominated atmospheres of $T_{\rm eff}\gtrsim 10,000\,\rm K$ 
require systematic increases due to thermohaline mixing,
often by several orders of magnitude.
These higher rates can be tested, especially via X-ray observations
\citep{Farihi18}.
Very thin hydrogen envelopes ($M_{\rm H}\lesssim 10^{-8}\, M_\odot$)
are not considered in the models presented here, but are often poorly
constrained. Such thin envelopes could
impose a systematic effect on inferences by lessening the
impact of thermohaline mixing, with smaller inferred accretion rates
as the result.
We have also assumed bulk earth composition, so objects that are very
rich in Ca relative to other metals could also be revised downward. 

Models that include
significant corrections due to thermohaline mixing also show very
different surface abundance ratios than those that only include
diffusive sedimentation. Diffusion leads to a correction to observed
compositions due to different sedimentation timescales for each element,
while models where thermohaline mixing dominates over diffusion show
metal ratios that match the accreted material.

We do not expect the thermohaline instability to play as large of a
role on  WDs with helium atmospheres
due to their much thicker convective envelopes that distribute
accreted metals much deeper into the star.
Preliminary \mesa\ runs with helium atmosphere WD models indicate
that thermohaline mixing is unimportant for those with
$T_{\rm eff}\lesssim 15,000\,\rm K$ due to their
thick (${M_{\rm cvz}\gtrsim 10^{-6}\, M_\odot}$) convective envelopes.
Hotter He atmosphere WDs may require some accretion rate corrections
due to thermohaline mixing, but we expect the effect to remain modest.
As predicted by \cite{Deal13}, our results in
Figure~\ref{fig:thermo_mdot} appear to resolve the discrepancy
between inferred accretion rates in helium and hydrogen atmospheres
that is seen in the work of \cite{Farihi12}.

The high accretion rates implied by our calculations here may pose
problems for models that deliver accreted metals to the WD surface via
Poynting-Robertson drag \citep{Rafikov11}.
Models of runaway accretion events due to other
disk processes have been proposed to account for high inferred
accretion rates in helium atmosphere WDs \citep{Rafikov11b,Metzger12},
but it is unclear if these can account for the highest rates suggested
in Figure~\ref{fig:thermo_mdot}.

Further work is necessary to extend our \mesa\ models to cover the
entire range of $\log g$ relevant to all observed polluted WDs. Future
work will present this along with more details from \mesa\
models for surface convection zone masses and individual diffusion
timescales for many elements, as well as models for WDs with helium
atmospheres.
Surface mixing regions in polluted WDs may also be modified by
convective overshoot \citep{Tremblay15,Kupka18}, and \mesa\ models
have the potential for including this effect as well.

\acknowledgments
{\it Acknowledgments}:
We thank Boris G{\"a}nsicke for drawing our
attention to the potential importance of
thermohaline mixing for our pollution calculations,
and for providing helpful comments to improve this paper.
We thank Jay Farihi for providing accretion rates and effective
temperatures for the DBZ sample shown in
Figure~\ref{fig:thermo_mdot}.
We thank Tim Brandt for insightful discussions.
We thank Bill Paxton for continuous efforts that enable 
the broad use of \MESA.
This work was supported by the National 
Science Foundation through grants
PHY 17-148958 and ACI 16-63688.

\software{
  \mesa\ \citep{Paxton11,Paxton13,Paxton15,Paxton18},
  \texttt{Matplotlib} \citep{Matplotlib},
  \texttt{NumPy} \citep{Numpy},
  \texttt{SciPy} \citep{scipy},
  \texttt{MesaScript} \citep{Mesascript}
}

\bibliographystyle{yahapj}

\end{document}